\documentclass[authoryear,preprint,review,12pt]{elsarticle}

\newcommand{\Msun}{\mbox{$M_{\odot}$}}
\newcommand{\Rsun}{\mbox{$R_{\odot}$}}

\newcommand{\kms}{\mbox{km s$^{-1}$}}

\usepackage{graphics}

\journal{New Astronomy}

\begin{document}

\begin{frontmatter}



\title{The mass and radius of the M-dwarf companion in the double-lined eclipsing binary T-Cyg1-01385}


\author[OCakirli]{\"{O}m\"{u}r \c{C}ak{\i}rl{\i}\corref{cor1}}
\ead{omur.cakirli@gmail.com, Tel:+90 (232) 3111740, Fax:+90(232) 3731403}
\cortext[cor1]{Corresponding author}

\address[OCakirli]{Ege University, Science Faculty, Astronomy and Space Sciences Dept., 35100 Bornova, \.{I}zmir, Turkey.}
\author[OCakirli]{Cafer Ibanoglu} 
\author[OCakirli]{E. Sipahi}

\begin{abstract}
We observed spectroscopically the eclipsing binary system T-Cyg1-01385 in order to determine physical properties of the components.
The double-lined nature of the system is revealed for the first time and the radial velocities are obtained for both stars. We have derived masses, radii and luminosities for both components. Analyses of the radial velocities and the KeplerCam and the T$r$ES light curves yielded masses of M$_1$=1.059$\pm$0.032 \Msun ~and M$_2$=0.342$\pm$0.017 {\Msun} and radii of R$_1$=1.989$\pm$0.022 {\Rsun} and R$_2$=0.457$\pm$0.013 {\Rsun}. Locations of the low-mass companion in the mass-radius and mass-effective temperature planes and comparison with the other low-mass stars show that the secondary star appears just at the transition from partially to fully convective interiors for the M dwarfs. When compared to stellar evolution models, the luminosities and effective temperatures of the components are consistent with Z=0.004 and an age of about 6 Gyr. A distance to the system was calculated as d=355$\pm$7 pc using the BV and JHK magnitudes.      
\end{abstract}

\end{frontmatter}



\section{Introduction}
The lower-mass stars of the main-sequence of the Hertzsprung-Russell diagram are known as M-dwarfs. They constitute
the majority of stars in the solar neighbourhood. They are intrinsically faint since they are cooler and smaller than the other stars .
Because of their faintness photometric and spectroscopic observations could be obtained for a limited number M-dwarfs.  \citet{Chabrier}
suggests that at least 70 per cent of all stars in the sphere with a radius of 10 pc about the Sun are M-dwarfs. Therefore, detailed studies of these
faint but numerous low-mass stars are critical importance to evaluate a model of the galactic evolution and present status. Moreover the M-dwarfs
are particularly needed in the understanding of evolution of main-sequence stars towards the lower-mass regime. The evolutionary calculations of low-mass stars allow us to define not only the transition from partly to fully convection but also to define a limit between stars and brown dwarfs. 
As pointed out by \citet{Shulyak} around spectral type M3.5 stars become fully convective and thus the dynamo mechanism must be different in cooler stars 
because they do not posses tachocline  layer with strong differential rotation. However, \citet{Reiners07} have concluded from the measurements of Stokes parameter about tens of stars spanning the whole M-dwarfs that no significant change in the average magnetic field strengths occurs when stars become fully convective. Later on \citet{Reiners09} could reveal the rotation- magnetic field relation in which magnetic field strengths increase towards short rotation periods. 

Low-mass M-dwarfs are generally faint objects because of their small radii and low temperatures. Due to their faintness the M-dwarfs have limited the number of high-resolution, high signal-to-noise spectroscopic studies. In addition, the optical spectrum of these cool stars is mostly covered by molecular bands which hide and blend the atomic lines used in spectral analysis. These dominant molecular bands make it difficult to measure the atomic line strengths which are needed for metallicity determination.  The physical parameters,  such as mass, radius and luminosity as well as age and rotational period could only be derived if a M-dwarf is a member of a close binary or a multiple system. In a binary or a multiple system the components are most likely coeval and their spin axes are perpendicular to the
orbital plane.           

Numerous photometric observations of binary stars were recently gathered by the surveys of like the T$r$ES\,\citep{Alonso}, NSVS\,\citep{woz}, SuperWASP\,\citep{Christian06},  Kepler\,\citep{bor}, etc. Despite the main aim of these surveys is to search gamma-ray bursters, and especially extra-solar planetary transits many binary systems are discovered and light curves of many systems could be obtained. Therefore, the tremendous photometric datasets containing unknown or little known binary systems are presented for the use of astronomers.

Light variability of T-Cyg1-01385 was announced by \citet{Devor08} and \citet{Devor2008} in the list of 773 eclipsing binaries found in the Trans-Atlantic Exoplanet Survey. It was classified as an "ambiguous" binary in their list. Mass and radius of the components were estimated for the first time by \citet{Fernandez09} combining the spectroscopic orbital elements obtained from the primary star's radial velocities  with a high-precision transit light curve obtained by the KeplerCam. The results obtained up-to-date point out that the fainter component of T-Cyg1-01385 is so close to the fully convective M-dwarfs. Therefore we planned new spectroscopic observations of the system to refine the masses, radii and effective temperatures of both stars.

\begin{table}
\scriptsize
\caption{Catalog information for T-Cyg1-01385.}
\setlength{\tabcolsep}{0.8pt}
\begin{tabular}{lcr} 
\hline
Source Catalog & Parameter &Value\\
\hline
2MASS$^{a}$					    & $\alpha$ (J2000)      				& 20$^h$15$^m$ 21.94$^s$\\	
2MASS 								& $\delta$ (J2000)      					& +48${\circ}$17$^{\prime}$14.14$^{\prime\prime}$\\
PPMX$^{b}$       							& $V$ mag 									& 10.92 $\pm$ 0.08\\
GSC2.3$^{c}$						& $B$ mag 									& 11.68 $\pm$ 0.08\\
GSC2.3								& $V$ mag 									& 11.02 $\pm$ 0.07\\
TYCHO$^{d}$						& $B_T$ mag 								& 11.68 $\pm$ 0.08\\
TYCHO								& $V_T$ mag 								& 11.02 $\pm$ 0.07\\		
2MASS  								& $J$ mag  									& 9.834 $\pm$ 0.026\\
2MASS 								& $H$ mag         							& 9.593 $\pm$ 0.031\\
2MASS 								& $K_s$ mag       						& 9.513 $\pm$ 0.021\\
NOMAD$^{e}$						& $B$ mag									& 11.520 $\pm$ 0.118\\	
NOMAD								& $V$ mag									& 10.956 $\pm$ 0.121\\
NOMAD								& $R$ mag									& 10.580 $\pm$ 0.150\\		
CMC14$^{f}$						& $r'$ mag 									& 10.892 $\pm$ 0.185\\
ASCC-2.5V3$^g$					&$B$ mag							    	& 11.565 $\pm$ 0.093\\
ASCC-2.5V3						&$V$ mag							    	& 10.920 $\pm$ 0.082\\
ACT2000.2$^h$					&$B$ mag							    	& 11.680 $\pm$ 0.013\\
ACT2000.2							&$V$ mag									& 11.016 $\pm$ 0.091\\
TASS Mark IV$^i$				&$V$ mag						        	& 11.103 $\pm$ 0.065\\
TASS Mark IV						&$I$ mag									    & 10.363 $\pm$ 0.049\\
UCAC4$^j$							&$B$ mag									&  11.741$\pm$0.011 \\
UCAC4								&$V$ mag									&  10.997$\pm$0.010 \\
2MASS 								& Identification								&2MASS20152193+4817141\\
UCAC4 								& Identification								&692-074290\\
TASS Mark IV					    & Identification								&  1937252 \\
TYCHO 								& Identification								& TYC3576-2035-1 \\
GSC2.3 								& Identification								& GSC\,03576\,02035\\
USNO-B1.0 						& Identification								&  1382-0386191 \\ 							
PPMX 							     	& Identification								&201521.9+481714   \\ 	
T$r$ES$^k$ 						    & Identification								& T-Cyg1-01385  \\ 	
SB9 $^l$							& Identification								& 3018   \\ 	
SWASP $^m$						& Identification								& 1SWASP\,J201521.94+481714.1   \\ 	
UCAC4				    & $\mu_\alpha$, $\mu_\delta$ (${\rm mas\,yr^{-1}}$)  		&-9.2$\pm$0.6, -22.4$\pm$1.2 \\
TYCHO				    & $\mu_\alpha$, $\mu_\delta$ (${\rm mas\,yr^{-1}}$)  		&-8.4$\pm$3.6, -23.4$\pm$3.1 \\
ASCC-2.5V3			& $\mu_\alpha$, $\mu_\delta$ (${\rm mas\,yr^{-1}}$)  		&-8.77$\pm$2.92, -20.37$\pm$2.51\\
USNO-B1.0			& $\mu_\alpha$, $\mu_\delta$ (${\rm mas\,yr^{-1}}$)  			&-10$\pm$1, -20$\pm$1 \\
PPMX						& $\mu_\alpha$, $\mu_\delta$ (${\rm mas\,yr^{-1}}$)  		&-7.8$\pm$1.6, -19.5$\pm$1.8 \\
\hline \end{tabular} \\
{\rm $^{a}${\em 2MASS} Two Micron All Sky Survey Catalog \citep{sku2006}}, 
{\rm $^{b}${\em PPMX} Position and Proper Motions Catalog,  \citep{roeser08}},
{\rm $^{c}${\em GSC2.3} Guide Star Catalog, version 2.3.2 \citep{mo99}}, 
{\rm $^{d}${\em TYCHO} Tycho Catalog,  \citep{esa}},
{\rm $^{e}${\em NOMAD} NOMAD Catalog,  \citep{zac05}},
{\rm $^{f}${\em CMC14} Carlsberg Meridian Catalog 14 \citep{car}}, 
{\rm $^{g}${\em ASCC-2.5V3 } All-Sky Compiled Catalog of 2.5 Millon Stars,  \citep{karco}}, 
{\rm $^{h}${\em ACT}  Astrographic Catalog,  \citep{act}}, 
{\rm $^{i}${\em TASS} The Amateur Sky Survey (TASS) Catalog,  \citep{drog}}, 
{\rm $^{j}${\em UCAC}  High density, Highly Accurate, Astrometric Catalog,  \citep{ucac}}, 
{\rm $^{k}${\em T$r$ES} Trans Eclipsing Binary Catalog,  \citep{Devor2008}}, 
{\rm $^{l}${\em SB}  9$^{th}$ Catalog of Spectroscopic Binary Orbits,  \citep{sb}}.
{\rm $^{m}${\em SuperWASP} \citep{Christian06} }.
\end{table}

\section{Period determination}

The catalog information for T-Cyg1-01385 was  given in Table\,1. First we collected the times for mid-light minimum obtained by various automatic and robotized telescopes and surveys. These times of minima are given in Table\,2 as averaged for the filters used. A linear least-squares fit 
to the data yields the following ephemeris  
\[ {\rm Min\,I} = {\rm HJD}\ 2453926.7896 (3) + 6^d.558662 (31) \times E \] 
where the bracketed quantity is the uncertainty in the last digit(s) of the preceding number. All 
uncertainties quoted in this work are standard errors. The residuals of the fit are plotted in 
Fig.\,1 and show no indication of any form of period change in about ten years. 

\begin{figure}
\includegraphics[width=9cm,angle=0]{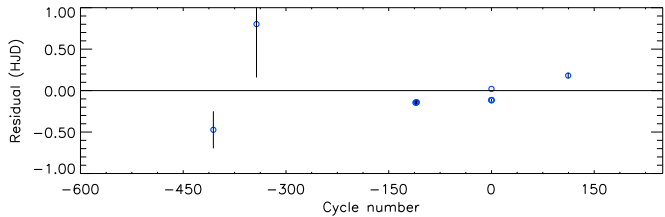}
\caption{The residuals between the observed times of minimum 
light and computed with the new ephemeris.} \end{figure}

\begin{table}
\scriptsize
\caption{Times of minimum light for T-Cyg1-01385. The O-C values refer to the difference between the 
observed and calculated times of mid-eclipse.}
\label{O-C values}
\begin{tabular}{cccc}
\hline
  Minimum time  & Cycle number   & O-C & Ref.  \\
   (HJD-2\,400\,000)            & &     & \\
  \hline
51263.5000	&	-406.0	&	0.0983  &1	\\
53198.6330	&	-111.0	&	-0.0041 &2\\
53205.1940	&	-110.0	&	-0.0032 &2	\\
53208.4730	&	-109.5	&	-0.0043 &2	\\
53211.7550	&	-109.0	&	-0.0023 &2	\\
53926.6737	&	0.0	&	-0.1367	&3\\
53926.6743	&	0.0	&	-0.1361	&3\\
53926.6755	&	0.0	&	-0.1349	&3\\
53926.6761	&	0.0	&	-0.1343	&3\\
53926.8104	&	0.0	&	0.0000	&3\\
54661.5413 &112.0& -0.00126&4\\
\hline
\end{tabular}
\begin{list}{}{}
\item[Ref:]{\small (1)  \citep{car}, (2) \citet{Devor08}, (3) \citet{Fernandez09}, (4) \citet{Christian06}}
\end{list}
\end{table}

\section{Echelle Spectroscopy}
Optical spectroscopic observations of the T-Cyg1-01385 were obtained with the RTT150\footnote{http://tug.tug.tubitak.gov.tr/rtt150\_tfosc.php} 
1.5 m telescope at TUG\footnote{T\"{U}B{\.I}TAK National Observatory (Turkey)} using the 
$R$ $\sim$ 7\,000 Echelle spectrograph that covers 400\,nm $\le$ $\lambda$ $\le$ 900\,nm. These 
observations were used to resolve the components of T-Cyg1-01385 and get individual radial 
velocity measurements for each star in the system. Eleven spectra were obtained using 60 minute 
integrations on 8\,nights in July, 2010 with typical signal-to-noise ratios of $\sim$120 at 6563 \AA.

The spectra were processed in the standard way for cross-dispersed Echelle spectra, using 
{\sc fxcor} package in IRAF. The routine processes the data using biases and halogen lamp 
observations taken at the beginning of the night, median combines three individual images while 
performing cosmic ray rejection, extracts the individual orders from the combined image and 
performs the wavelength solution on each order using a FeAr\,arc lamp taken either before or 
after each set of science exposures.

A total of 11 orders are used each night to derive radial velocities via cross-correlation with a 
standard template.  We use the bright $\iota$\,Psc (F7\,V) and 50\,Ser (F0\,V) as the heliocentric 
radial velocity standard stars. Each spectral order is cross-correlated separately, then an iterative 
3-$\sigma$ clipping is performed prior to performing a weighted average to obtain a final radial 
velocity measurement for each night. The components are identified each night via the peak and 
width of each feature in the cross correlated functions. The typical radial velocity precisions ranged from 
2 to 13 $\rm{km~s^{-1}}$ for the various components.

\subsection{Radial velocities}
To derive the radial 
velocities, the eleven spectra obtained for the system are cross-correlated 
against the template spectra of standard stars on an order-by-order basis using the {\sc fxcor} 
package in IRAF \citep{Simkin}. 

The majority of the spectra showed two distinct cross-correlation peaks in the quadratures, one 
for each component of the binary. Thus, both peaks are fitted independently
with a $Gaussian$ profile to measure the velocities and their errors for the individual components. If the 
two peaks appear blended, a double Gaussian was applied to the combined profile using {\it de-blend} 
function in the task. For each of the eleven observations we then determined a weighted-average radial 
velocity for each star from all orders without significant contamination by telluric absorption features. Here 
we used as weights the inverse of the variance of the radial velocity measurements in each order, as 
reported by {\sc fxcor}.

The heliocentric radial velocities for the primary (V$_p$) and the secondary (V$_s$) components are 
listed in Table\,3 , along with the dates of observations and the corresponding orbital phases computed 
with the new ephemeris given in the previous section. The velocities in this table have been corrected to 
the heliocentric reference system by adopting a radial velocity value for the template stars. The 
radial velocities are plotted against the orbital phase in Fig.\,2. The radial velocities of the more massive star measured by  
\citet{Fernandez09} are also plotted as circles in the same figure. There is no systematic difference between the measurements.

We analysed the radial velocities for the initial orbital parameters using the {\sc RVSIM} 
software program \citep{kane}. Figure\,2 shows the best-fit orbital solution to the radial velocity 
data. The results of the analysis are as follow: 
$\gamma$= -9$\pm$1 \kms, $K_1$=31$\pm$3 and $K_2$=96$\pm$4 \kms with circular orbit. Using these 
values we estimate the projected orbital semi-major axis and mass ratio 
as: $a$sin$i$=16.46$\pm$0.18 \Rsun~ and $q=\frac{M_2}{M_1}$=0.323$\pm$0.020.

\begin{figure}
\includegraphics[width=10cm,angle=0]{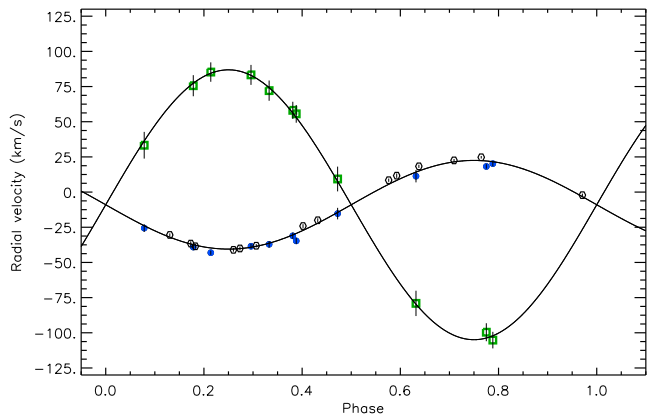}\\
\caption{Radial velocities for the components folded on a period of  6.5587 days. Symbols with error 
bars (error bars are masked by the symbol size in some cases) show the radial velocity 
measurements for the components of the system (primary: filled circles, secondary: open 
squares). The velocities measured by \citet{Fernandez09} are shown by empty circles.} \end{figure}

\begin{table}
\scriptsize
\centering
\begin{minipage}{85mm}
\caption{Heliocentric radial velocities of T-Cyg1-01385. The columns give the heliocentric 
Julian date, the orbital phase (according to the ephemeris in \S 2), the radial velocities of 
the two components with the corresponding standard deviations.}

\begin{tabular}{@{}ccccccccc@{}c}
\hline
HJD 2400000+ & Phase & \multicolumn{2}{c}{Star 1 }& \multicolumn{2}{c}{Star 2 } 	\\
             &       & $V_p$                      & $\sigma$                    & $V_s$   	& $\sigma$	\\
\hline
55390.5404 &	0.1783 & -39.1 &  1.6	&   75.6   &  7.5  \\
55391.3117 &	0.2959 & -38.5 &  1.8	&   83.3   &  7.1  \\
55391.5538 &	0.3328 & -37.1 &  2.3	&   72.1   &  7.2  \\
55392.4691 &	0.4723 & -15.3 &  4.1	&     9.3   &  8.8  \\
55393.5155 &	0.6319 &  11.3 &  4.4	&  -79.1   &  9.0  \\
55394.4550 &	0.7751 &  18.2 &  1.6	&  -99.6   &  6.4  \\
55394.5423 &	0.7884 &  20.2 &  1.8	& -105.2  &  5.9  \\
55396.4442 &	0.0784 & -25.6 &  2.7	&    32.3  &  13.5 \\
55397.3337 &	0.2140 & -43.1 &  1.5	&    85.4  &  6.9  \\
55398.4284 &	0.3810 & -31.1 &  2.2	&    58.1  &  6.1  \\
55398.4746 &	0.3880 & -34.7 &  2.2	&    55.6  &  6.3  \\

\hline \\
\end{tabular}
\end{minipage}
\end{table}

\section{Light curves and their analyses}
Photometric observations of T-cyg1-01385 were obtained by automatic and robotised telescopes. The first complete 
light curve was obtained by the T$r$ES wide-angle transiting planet survey \citep{Alonso}. Additional photometric 
data, especially in the primary eclipse, were obtained by the NSVS \citep{woz}, SuperWASP\,\citep{Christian06}, 
and  $Kepler$Cam\citep{bor}. Since the data gathered by the SuperWASP have too large scatters we do not 
include them into the analysis for the orbital parameters. In Fig.\,3 we plotted the T$r$ES data against orbital 
phase. The observations obtained by the T$r$ES and the $Kepler$Cam within the primary eclipse are shown 
in Fig.4.  

\begin{figure*}
\center
\includegraphics[width=16cm,angle=0]{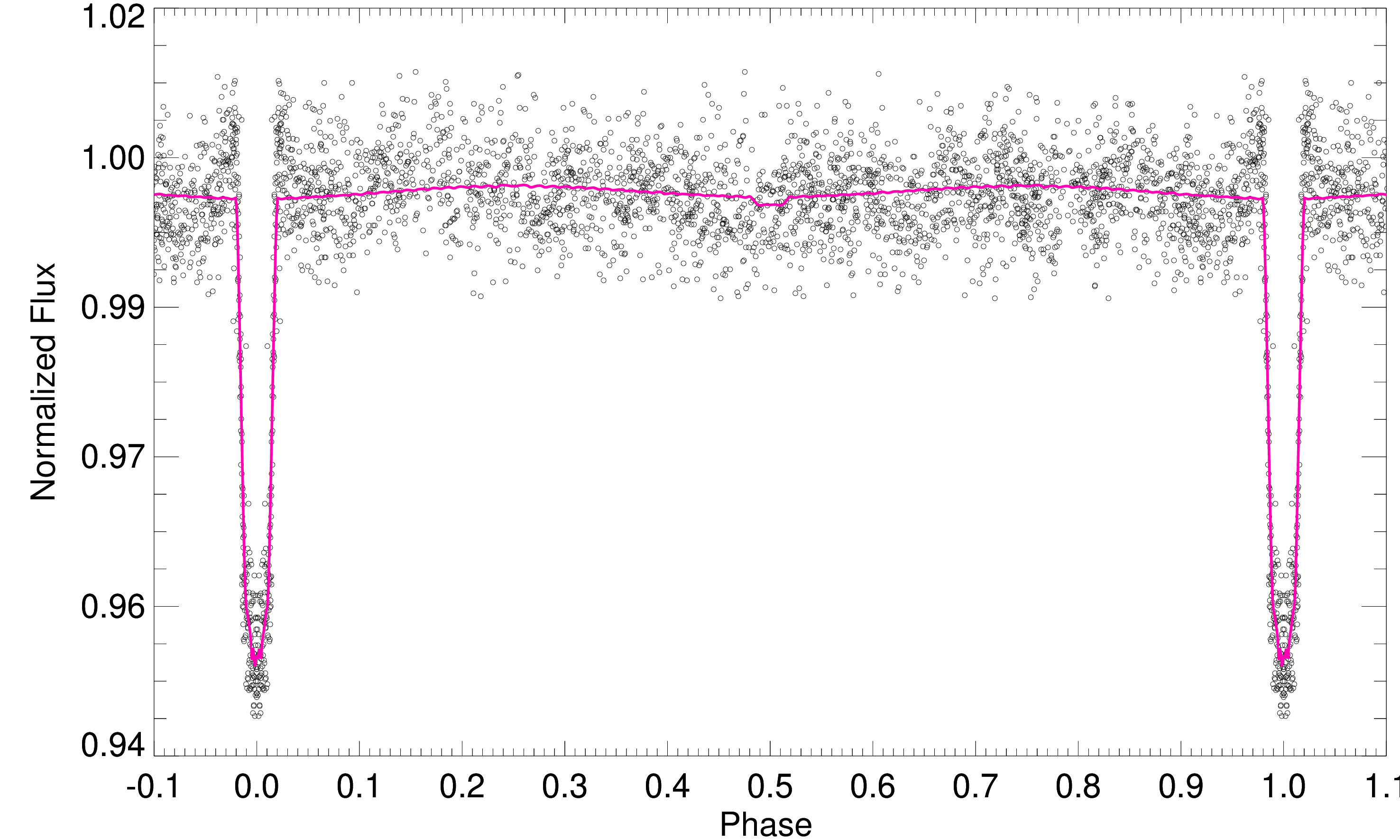}
\caption{Unbinned phase-folded light curve of T-Cyg1-01385 obtained by the T$r$ES in the $Sloan\,r$-passband 
and the best-fit model. The very shallow secondary eclipse is detectable at around phase 0.5. } \end{figure*}

We may constrain the effective temperature and spectral type of the primary star using the B$_T$,  V$_T$, J, H, 
and K  magnitudes which are already given in Table\,1. We have derived V=10.96, B-V=0.56, J-H=0.241 and 
H-K=0.08 mag. Comparing the color indices with color-spectral type calibrations given by \citet{drill00} and 
\citet{tok00} we estimated a spectral type of F8 sub-giant for the primary star. Thus, an effective temperature 
of  $T_{\rm eff} = 6\,250 \pm 100 $K and a color excess of 0.04 mag are estimated.

\begin{figure}
\includegraphics[width=8cm,angle=0]{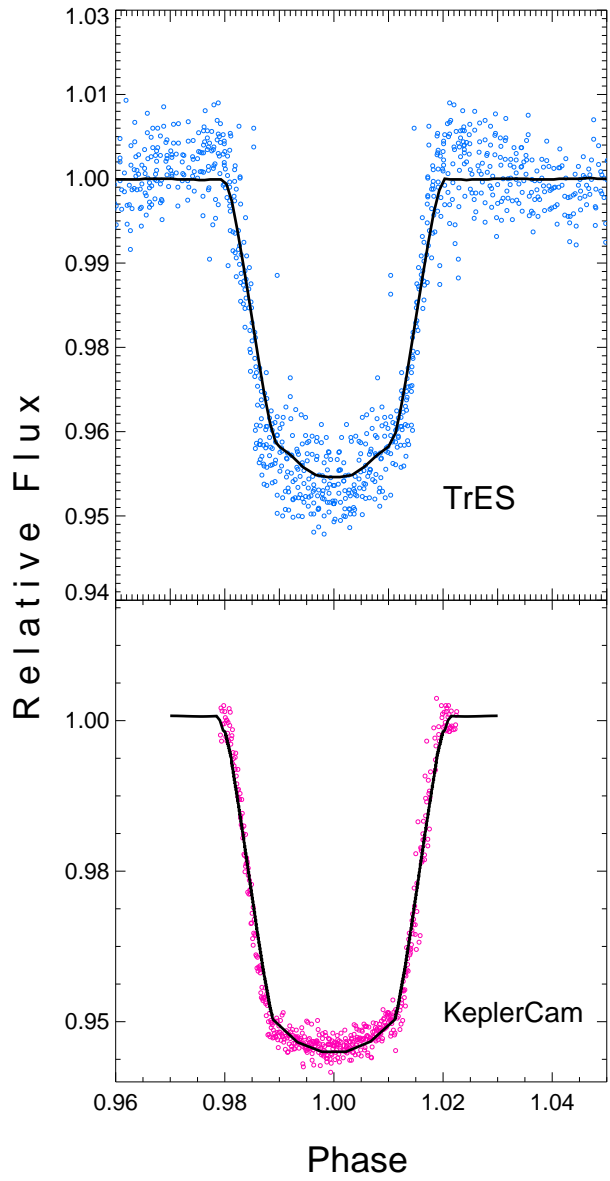}
\caption{Eclipse light curves of T-Cyg1-01385 obtained by the T$r$ES at the $Sloan\,r$-passband and the $Kepler$Cam at the z-passband. The continuous lines show the best fits.  } \end{figure}

We used the most recent version of the eclipsing binary light curve modeling algorithm of \citet{wil}, as implemented 
in the {\sc phoebe} code of Pr{\v s}a \& Zwitter (2005). In order to obtain the physical parameters of the component 
stars we, first analysed the T$r$ES data. The code needs some input parameters, which depend upon the physical 
structures of the component stars. The values of these parameters can be estimated from global stellar 
properties. Therefore, we adopted the linear limb-darkening coefficients from \citet{vanham93}, the bolometric 
albedos from \citet{lucy} and the gravity brightening coefficients as 0.32 for both components. The 
rotational velocities of the components are assumed to be synchronous with the orbital one.

The adjustable parameters in the light curves fitting were the orbital inclination, the surface potentials 
of the two stars, the effective temperature of the secondary, and the color-dependent luminosity of the hotter 
star, the zero-epoch offset, semi-major axis of the orbit, the mass-ratio and the systemic velocity. A detached 
configuration (Mode 2) with coupling between luminosity and temperature was used for solution. The iterations 
were carried out automatically until convergence, and a solution was defined as the set of parameters for which 
the differential corrections were smaller than the probable errors. Our final results are listed in Table\,4. The 
uncertainties assigned to the adjusted parameters are the internal errors provided directly by the code.  The 
computed light curve corresponding to the simultaneous light-velocity solution is compared with the 
observations in the Fig.\,3 and 4.

\begin{table}
\scriptsize
\caption{Results of the simultaneous analyses of the T$r$ES and the $Kepler$Cam light curves for T-Cyg1-01385.}
\begin{tabular}{lr}
\hline
Parameters & Adopted  \\
\hline	
$i^{o}$			               			 					&86.4$\pm$0.02				\\
T$_{eff_1}$ (K)												&6\,250[Fix]						\\
T$_{eff_2}$ (K)												&2\,940$\pm$120			\\
$\Omega_1$													&8.609$\pm$0.023			\\
$\Omega_2$													&12.989$\pm$0.030		\\
$r_1$																&0.1206$\pm$0.0003		\\
$r_2$																&0.0272$\pm$0.0001		\\
$\frac{L_{1}}{(L_{1}+L_{2})}$ (T$r$ES-r) 	&0.9993$\pm$0.0005			\\
$\frac{L_{1}}{(L_{1}+L_{2})}$ ($Kepler$-z)&0.9996$\pm$0.0004			\\
$\chi^2$															&0.051								\\				
\hline
\end{tabular}
\end{table}

The fundamental stellar parameters for the components such as masses, radii, luminosities were calculated 
and listed in Table\,5 together with their formal standard deviations. The standard deviations of the parameters
have been determined by JKTABSDIM\footnote{This can be obtained from http://http://www.astro.keele.ac.uk/$\sim$jkt/codes.html} code, which calculates distance and other physical parameters using several different sources of bolometric 
corrections \citep{south}. The mass for the primary of $M_A$ = 1.06 $\pm$ 0.03M $_{\odot}$  and secondary 
of $M_B$ = 0.34 $\pm$ 0.02M$_{\odot}$ are consisting of an evolved late F-star and mid M-dwarf \citep{drill00}.

\begin{table}
\scriptsize
 \setlength{\tabcolsep}{2.5pt} 
  \caption{Fundamental parameters of T-Cyg1-01385}
  \label{parameters}
  \begin{tabular}{lcc}
  \hline
   Parameter 																& Primary	&	Secondary										\\
   \hline
   Spectral Type																	& F8($\pm$2)IV-V  		& M4($\pm$1)V    			\\
   $a$ (R$_{\odot}$)																&\multicolumn{2}{c}{16.49$\pm$0.18}			\\
   $V_{\gamma}$ (km s$^{-1}$)											&\multicolumn{2}{c}{-9$\pm$1}					\\
   $i$ ($^{\circ}$)																	&\multicolumn{2}{c}{86.36$\pm$0.02}				\\
   $q$																					&\multicolumn{2}{c}{0.323$\pm$0.020}		\\
   Mass (M$_{\odot}$) 															& 1.059$\pm$0.032 	& 0.342$\pm$0.017			\\
   Radius (R$_{\odot}$) 														& 1.989$\pm$0.022 	& 0.457$\pm$0.013			\\   
   $T_{eff}$ (K)																		& 6\,250$\pm$100	& 2\,940$\pm$120		\\
   $\log~(L/L_{\odot})$															& 0.736$\pm$0.035		& 2.039$\pm$0.083		\\
   $\log~g$ ($cgs$) 																& 3.866$\pm$0.005 	& 4.652$\pm$0.026			\\
   $(vsin~i)_{calc.}$ (km s$^{-1}$)										& 15$\pm$1				& 4$\pm$1					\\   
   $d$ (pc)																			& \multicolumn{2}{c}{355$\pm$7}				\\
\hline  
  \end{tabular}
\end{table}

There is no measured trigonometric parallax available for the system. From the B- and V-passband 
measurements of Tycho and the JHK magnitudes given in the 2MASS catalog we estimated an 
interstellar reddening of $E(B-V) \simeq 0.04$ mag. Then we estimated an average distance to the 
system as 355$\pm$7 pc.

\section{Comparison with models and other low-mass stars}
Using the radii and effective temperatures we computed the luminosities of the components as 
$L_1$ = 5.4 $\pm$ 0.4L $_{\odot}$  and  $L_2$ = 0.009 $\pm$ 0.002L $_{\odot}$ for the primary 
and secondary, respectively. In Fig.\,5 we compare the positions of the stars in the $L-T_{eff}$ 
diagram. The isochrones for 5, 6, and 7 Gyr obtained by $Y^{2}$ models (\citep{Yi01}, 
\citep{Demarque04}) for Z=0.004 are also plotted. The primary star appears to an 
evolved F-star with an age of about 6 Gyr. This comparison with the theoretical models 
indicates that the primary star should have poor metal abundance. 

\begin{figure}
\center
\includegraphics[width=14cm,angle=0]{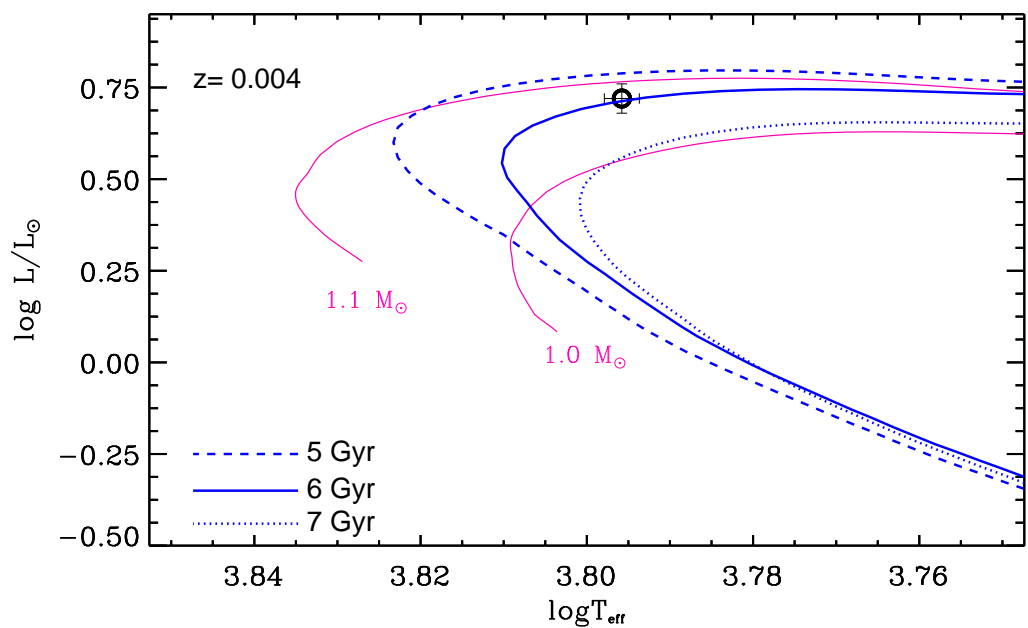}
\caption{The primary component of T-Cyg1-01385 compared with Y$^2$ models for Z=0.004. For comparison 
the evolutionary tracks for 1 and 1.1M$_{\odot}$ and isochrones for 5, 6 and 7 Gyr are shown. } \end{figure}

For the first time \citet{Ribas03} and \citet{Ribas06} called attention about the significant difference of low-mass 
stars' radii between measured and predicted by the theory. His comparison for a limited stars with masses and 
radii determined with an accuracy better than 3 percent revealed that the observed radii are systematically larger 
than those estimated by theoretical models. In contrary, their effective temperatures are significantly cooler than in 
the models. Later on, \citet{Cakirli10} collected the available masses, radii and effective temperatures of the 
low-mass stars obtained up to at that date and compared with those calculated by the stellar evolution models. 
They have also compared radii and effective temperatures of 21 low-mass stars in the mass-radius and 
mass-effective temperature diagrams. They conclude that stars below 0.3 M$_{\odot}$ have radii and 
effective temperatures which are consistent with those of models. However, larger mass stars, above 
0.3 M$_{\odot}$, begin to deviate from the theoretical predictions in the $M-R$ diagram. The observed 
radii are significantly larger than that of models. In contrary, their observed effective temperatures are 
significantly lower than those models, i.e. they appear as below-shifted from the theoretical  $M-T_{eff}$ 
relation. The discrepancies in radii and effective temperatures could neither be explained by changing the 
ratio of mixing-length to pressure scale height nor metallicity \citep{Demory09}. As an alternative explanation 
the magnetic activity, which is responsible for the observed larger radii but cooler effective temperatures, was 
adopted by \citet{Mullan01}, \citet{Torres06}, \citet{Ribas06} and \citet{Cakirli13}.
Due to high magnetic activity in fast-rotating dwarfs, because of spin-orbit synchronization, their surfaces are 
covered by dark spots.  Faster rotation enhances solar-type activity which results in larger radius and lower 
effective temperature.

Fig.\,6 shows locations of the low-mass stars in the  in $M-R$ and $M-T_{eff}$ planes. The low-mass companion 
of T-Cyg1-01385 appears to have a radius about 26 per cent larger and effective temperature about 18 per cent 
cooler than the models \citep{Baraffe98}. In a very recent paper \citet{Cakirli13} called attention to the stars having 
masses about 0.34 M$_{\odot}$. The empirical $M-R$ diagram shows that the stars with masses lower than about 
0.3 M$_{\odot}$ do not deviate from that predicted by the models. However, largest deviations both in $M-R$ and 
$M-T_{eff}$ occur just at the mass higher than 0.34 M$_{\odot}$. This mass may be taken as transition from partly 
convective atmospheres to the fully convective stars.

\begin{figure*}
\center
\includegraphics[width=14cm,angle=0]{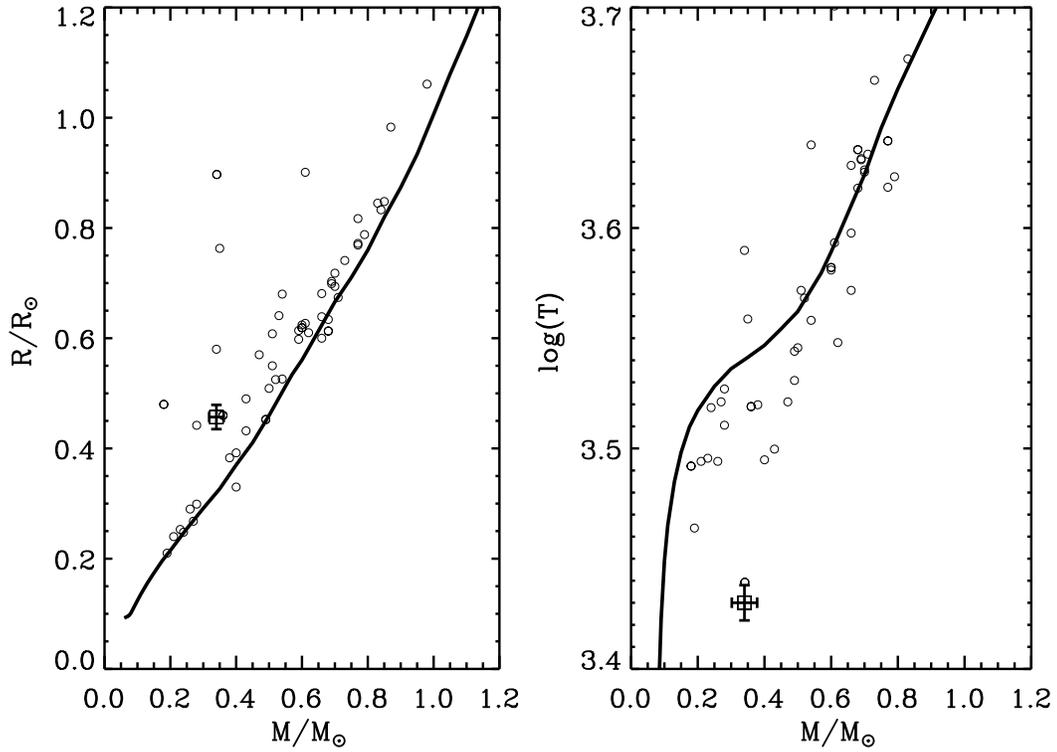}
\caption{Mass-radius (left) amd mass-temperature (right) diagrams of the stars between 0.2 to 1.0 M$_{\odot}$
in double-lined, detached eclipsing binaries. The continuous line corresponds to zero-age main-sequence adopted 
from \citet{Baraffe98}  models. The radius and effective temperature for the less massive companion of 
T-Cyg1-01385 are plotted as the open squares with standard deviations. } \end{figure*}

\section{Conclusions}
Spectroscopic observations of T-Cyg1-01385 are obtained and the radial velocity curves of both components are 
revealed for the first time. Analysis of the radial velocities and the T$r$ES at the $Sloan\,r$-passband and the 
$Kepler$Cam at the z-passband light curves yielded the physical parameters of the components. Our analysis
indicates that the eclipsing binary consists of an F8 subgiant and an M4 dwarf. {\bf The radius of the less massive 
M-dwarf is larger about 26 percent but the effective temperature 18 percent cooler than those estimated from the 
models.} The radii and effective temperatures of low-mass stars obtained so far have been compared with each 
other and also with the models both in the $M-R$ and $M-T_{eff}$ panels. {\bf The less massive secondary star with 
a mass of 0.34M$_{\odot}$ appears to be just in the boundary between partially convection and fully convection.}

\section*{Acknowledgments}
We thank to T\"{U}B{\.I}TAK National Observatory (TUG) for a partial support in using RTT150, T100 and T60 
telescopes with project numbers 10ARTT150-483-0, 11ARTT150-123-0, 10CT100-101, 12BRTT150-338-1 and 
10CT60-72. 
This study supported by the Turkish Scientific and Technology Council under project number 112T016 and 112T263.
We also thank to the staff of the Bak{\i}rl{\i}tepe observing station for their warm hospitality. 
The following internet-based resources were used in research for this paper: the NASA Astrophysics Data 
System; the SIMBAD database operated at CDS, Strasbourg, France; T\"{U}B\.{I}TAK ULAKB{\.I}M S\"{u}reli 
Yay{\i}nlar Katalo\v{g}u-TURKEY; and the ar$\chi$iv scientific paper preprint service operated by Cornell 
University.

\end{document}